\input harvmac
\input epsf

\def\al{\alpha}
\def\at{\tilde{\alpha}}
\def\ap{\alpha'}


\Title{}{\vbox{\centerline{A Matrix Model for Misner Universe }
\centerline{and Closed String Tachyons}}}

 \centerline{Jian-Huang
She\footnote{}{Emails: jhshe@itp.ac.cn}}

\medskip
\centerline{\it  Institute of Theoretical Physics, Chinese Academy of Science, }
 \centerline{\it P.O.Box 2735, Beijing 100080, P.R. China}
\medskip

\medskip
\centerline{\it  Graduate School of the Chinese Academy of Sciences, Beijing 100080, P.R. China}

\medskip

We use D-instantons to probe the geometry of Misner universe, and
 calculate the world volume field theory action, which is of the 1+0 dimensional form
and highly
 non-local. Turning on closed string tachyons, we see from
the deformed moduli space of the D-instantons that the spacelike
singularity is removed and the region near the singularity becomes
a fuzzy cone, where space and time do not commute. When realized
cosmologically there can be controllable trans-planckian effects.
 And the infinite past is now causally connected with the infinite
future, thus also providing a model for big crunch/big bang
transition. In the spirit of IKKT matrix theory, we propose that
the D-instanton action here provides a holographic description for
Misner universe and time is generated dynamically. In addition we
show that winding string production from the vacua and instability
of D-branes have
 simple uniform interpretations in this second quantized formalism.

\Date{Sep. 2005}

\nref\orb{Lance J. Dixon, Jeffrey A. Harvey, C. Vafa, Edward Witten,
"Strings on orbifolds", Nucl.Phys.B261:678-686,1985; "Strings on orbifolds.2", Nucl.Phys.B274:285-314,1986.}

\nref\andi{A. Strominger, "Massless black holes and conifolds in string theory", Nucl. phys. B
451, 96 (1995), hep-th/9504090.}

\nref\enh{Clifford V. Johnson, Amanda W. Peet, Joseph Polchinski,
"Gauge Theory and the Excision of Repulson Singularities",  Phys.Rev. D61 (2000) 086001, hep-th/9911161.}

\nref\horowitz{Gary T. Horowitz, Alan R. Steif, "Singular string
solution with nonsigular initial data", Phys.Lett.B258:91-96,1991.}

\nref\niki{Nikita A. Nekrasov, "Milne universe, tachyons and quantum group", Surveys High Energ.Phys.17:115-124,2002,
e-Print Archive: hep-th/0203112.}

\nref\Berkz{Micha Berkooz, Ben Craps, David Kutasov, Govindan Rajesh,
"Comments on cosmological singularities in string theory", JHEP 0303:031,2003,
e-Print Archive: hep-th/0212215.}

\nref\Berka{M. Berkooz, B. Pioline, "Strings in an electric field, and
the Milne Universe", JCAP 0311 (2003) 007, hep-th/0307280.}

\nref\Berkb{M. Berkooz, B. Pioline, M. Rozali, "Closed Strings in
 Misner Space: Cosmological Production of Winding Strings",
JCAP 0408 (2004) 004, hep-th/0405126. }

\nref\Berkc{M. Berkooz, B. Durin, B. Pioline, D. Reichmann,
"Closed Strings in Misner Space: Stringy Fuzziness with a Twist",
 JCAP 0410 (2004) 002, hep-th/0407216.}

 \nref\pio{Bruno Durin, Boris Pioline, "Closed strings in Misner
 space: a toy model for a Big Bounce ?", Proceedings of the NATO
 ASI and EC Summer School ``String Theory: from Gauge Interactions
 to Cosmology'', Cargese, France, June 7-19, 2004, hep-th/0501145.}

\nref\APS{A. Adams, J. Polchinski, E. Silverstein, "Don't panic! Closed string tachyons
in ALE space-times", JHEP 0110:029,2001,
e-Print Archive: hep-th/0108075}

\nref\min{Matthew Headrick, Shiraz Minwalla, Tadashi Takayanagi,
"Closed String Tachyon Condensation: An Overview",
Class.Quant.Grav. 21 (2004) S1539-S1565,  e-Print Archive:hep-th/0405064.}

\nref\sil{A. Adams, X. Liu, J. McGreevy, A. Saltman, E. Silverstein,
"Things Fall Apart: Topology Change from Winding Tachyons", hep-th/0502021.}

\nref\sm{John McGreevy, Eva Silverstein, "The Tachyon at the End of
the Universe",hep-th/0506130.}

\nref\DKPS{Michael R. Douglas, Daniel Kabat, Philippe Pouliot, Stephen H. Shenker,
"D-branes and Short Distances in String Theory", Nucl.Phys. B485 (1997) 85-127,
hep-th/9608024.}

\nref\Jap{Yasuaki Hikida, Rashmi R. Nayak, Kamal L. Panigrahi,
"D-branes in a Big Bang/Big Crunch Universe: Misner Space", hep-th/0508003.}

\nref\BFSS{T. Banks, W. Fischler, S.H. Shenker, L. Susskind,
"M Theory As A Matrix Model: A Conjecture", Phys.Rev. D55 (1997) 5112-5128, hep-th/9610043.}

\nref\IKKT{N. Ishibashi, H. Kawai, Y. Kitazawa, A. Tsuchiya, "A
Large-N Reduced Model as Superstring",  Nucl.Phys. B498 (1997)
467, hep-th/9612115.}

\nref\Berkooz{Micha Berkooz, Zohar Komargodski, Dori Reichmann,
Vadim Shpitalnik, "Flow of geometry and instantons on the null
orbifold", hep-th/0507067.}

\nref\Taylor{Washington Taylor, IV,"D-brane field theory on compact
spaces", Phys.Lett.B394:283-287,1997, e-Print Archive:
hep-th/9611042.}

\nref\simon{Jos茅 Figueroa-O'Farrill, Joan Sim贸n, "Generalised
supersymmetric fluxbranes", JHEP 0112 (2001) 011, hep-th/0110170.}

\nref\sethi{Akikazu Hashimoto, Savdeep Sethi, "Holography and string
dynamics in time dependent backgrouns",
Phys.Rev.Lett.89:261601,2002, e-Print Archive: hep-th/0208126. }

\nref\ohta{Joan Simón, "Null orbifolds in AdS, Time Dependence
and Holography", JHEP 0210 (2002) 036, hep-th/0208165;  Rong-Gen
Cai, Jian-Xin Lu, Nobuyoshi Ohta, "NCOS and D-branes in
Time-dependent Backgrounds", Phys.Lett. B551 (2003) 178-186,
hep-th/0210206. }

\nref\ooo{ Hirosi Ooguri, Kostas Skenderis, "On The Field Theory Limit Of D-Instantons",
JHEP 9811 (1998) 013, hep-th/9810128.}

\nref\shenker{Lukasz Fidkowski, Veronika Hubeny, Matthew Kleban, Stephen Shenker,
"The Black Hole Singularity in AdS/CFT", JHEP 0402 (2004) 014, hep-th/0306170.}

\nref\hong{Guido Festuccia, Hong Liu, "Excursions beyond the
horizon: Black hole singularities in Yang-Mills theories (I)",
hep-th/0506202.}

\nref\dm{Michael R. Douglas, Gregory Moore, "D-branes, Quivers, and
ALE Instantons", hep-th/9603167. }

\nref\hull{C.M. Hull, "Timelike T-Duality, de Sitter Space, Large $N$
Gauge Theories and Topological Field Theory", JHEP 9807 (1998) 021, hep-th/9806146;
"Duality and the Signature of Space-Time", JHEP 9811 (1998) 017, hep-th/9807127.}

\nref\sen{Ashoke Sen, "Tachyon Condensation on the Brane Antibrane
System", JHEP 9808 (1998) 012, hep-th/9805170;   "Rolling
Tachyon", JHEP 0204 (2002) 048, hep-th/0203211;  "Tachyon Dynamics
in Open String Theory", Int.J.Mod.Phys. A20 (2005) 5513-5656,
hep-th/0410103.}

\nref\stalk{Eva silverstein, "The tachyon at the end of the universe", talk
at string2005,
 http://www.fields.utoronto.ca/audio/05-06/strings/silverstein/.}

\nref\bang{Justin Khoury, Burt A. Ovrut, Nathan Seiberg, Paul J. Steinhardt, Neil Turok,
"From Big Crunch to Big Bang", Phys.Rev. D65 (2002) 086007, hep-th/0108187.}

\nref\li{Y. Yoneya, in "Wandering in the Fields", eds. K. Kawarabayashi, A. Ukawa (World Scientific, 1987),
p419;  Miao Li, Tamiaki Yoneya, "D-Particle Dynamics and The Space-Time Uncertainty Relation",
Phys.Rev.Lett. 78 (1997) 1219-1222, hep-th/9611072; Miao Li, Tamiaki Yoneya,
"Short-Distance Space-Time Structure and Black Holes in String Theory : A Short Review of the Present Status",
in the special issue of the Journal of Chaos, Solitons and Fractals on "Superstrings, M, F, S...Theory", hep-th/9806240;
 Tamiaki Yoneya, "String Theory and the Space-Time Uncertainty Principle", Prog.Theor.Phys. 103 (2000) 1081-1125,
hep-th/0004074.}

\nref\swnc{Nathan Seiberg, Edward Witten, "String Theory and Noncommutative Geometry",
JHEP 9909 (1999) 032, hep-th/9908142.}

\nref\huang{Chong-Sun Chu, Brian R. Greene, Gary Shiu,
"Remarks on Inflation and Noncommutative Geometry", Mod.Phys.Lett. A16 (2001) 2231-2240, hep-th/0011241;
Stephon Alexander, Robert Brandenberger, Joao Magueijo, "Non-Commutative Inflation",
Phys.Rev. D67 (2003) 081301, hep-th/0108190; Robert Brandenberger, Pei-Ming Ho,
"Noncommutative Spacetime, Stringy Spacetime Uncertainty Principle, and Density Fluctuations",
Phys.Rev. D66 (2002) 023517; AAPPS Bull. 12N1 (2002) 10-20, hep-th/0203119; Qing Guo Huang, Miao Li,
"CMB Power Spectrum from Noncommutative Spacetime", JHEP 0306 (2003) 014, hep-th/0304203;
Qing-Guo Huang, Miao Li, "Noncommutative Inflation and the CMB Multipoles", JCAP 0311 (2003) 001, astro-ph/0308458;
Qing-Guo Huang, Miao Li, "Power Spectra in Spacetime Noncommutative Inflation",
Nucl.Phys. B713 (2005) 219-234, astro-ph/0311378.
}

\nref\tseyy{A.A. Tseytlin, "On non-abelian generalisation of Born-Infeld action in string theory",
Nucl.Phys. B501 (1997) 41-52, hep-th/9701125.}

\nref\fdst{Miao Li, "Strings from IIB Matrices", Nucl.Phys.B499(1997)149-158,
 hep-th/9612222;
 I. Chepelev, Y. Makeenko, K. Zarembo,
"Properties of D-Branes in Matrix Model of IIB Superstring", Phys.Lett. B400 (1997) 43-51,
hep-th/9701151; Ansar Fayyazuddin, Douglas J. Smith, "P-brane solutions in IKKT IIB matrix theory",
Mod.Phys.Lett. A12 (1997) 1447-1454; hep-th/9701168; A. Fayyazuddin, Y. Makeenko, P. Olesen, D.J. Smith, K. Zarembo,
"Towards a Non-perturbative Formulation of IIB Superstrings by Matrix Models", Nucl.Phys. B499 (1997) 159-182, hep-th/9703038.}

\nref\tsey{I. Chepelev, A.A. Tseytlin, "Interactions of type IIB D-branes from D-instanton matrix model",
Nucl.Phys. B511 (1998) 629-646, hep-th/9705120.}

\nref\mvkms{John McGreevy, Herman Verlinde, "Strings from
Tachyons", JHEP 0312 (2003) 054, hep-th/0304224; Igor R. Klebanov,
Juan Maldacena, Nathan Seiberg, "D-brane Decay in Two-Dimensional
String Theory", JHEP 0307 (2003) 045, hep-th/0305159. }

\nref\bss{Tom Banks, Nathan Seiberg, Stephen Shenker, "Branes from Matrices",
Nucl.Phys. B490 (1997) 91-106, hep-th/9612157.}

\nref\mbb{Ben Craps, Savdeep Sethi, Erik Verlinde,"A Matrix Big
Bang", hep-th/0506180;  Miao Li, "A Class of Cosmological Matrix
Models", hep-th/0506260;  Miao Li, Wei Song, "Shock Waves and
Cosmological Matrix Models", hep-th/0507185; Sumit R. Das, Jeremy
Michelson, "pp Wave Big Bangs: Matrix Strings and Shrinking Fuzzy
Spheres", hep-th/0508068; Bin Chen, "The Time-dependent
Supersymmetric Configurations in M-theory and Matrix Models",
hep-th/0508191; Bin Chen, Ya-li He, Peng Zhang, " Exactly Solvable
Model of Superstring in Plane-wave Background with Linear Null
Dilaton", hep-th/0509113.}

\nref\hhh{Thomas Hertog, Gary T. Horowitz, "Holographic
Description of AdS Cosmologies", JHEP 0504 (2005) 005,
hep-th/0503071.}

\nref\yz{Haitang Yang, Barton Zwiebach, "Rolling Closed String
Tachyons and the Big Crunch", hep-th/0506076;  Haitang Yang,
Barton Zwiebach, "A Closed String Tachyon Vacuum ?",
hep-th/0506077.}

\nref\tai{Yasuaki Hikida, Ta-Sheng Tai, "D-instantons and Closed
String Tachyons in Misner Space", hep-th/0510129.}

\newsec{Introduction}
The resolution of spacelike singularities is one of the most
outstanding problems in the study of quantum gravity.
These singularities make appearance in many black holes and
cosmological models. Unfortunately it is very hard to get much
information about them in general situations. So in order
to make progress on this issue, more controllable toy models are
proposed, the simplest of which may be the two dimensional Misner
space, which can be defined as the quotient of two dimensional
Minkowski space by a boost transformation.

Nowadays string theory is widely regarded as the most
promising candidate for a quantum theory of gravity. And actually string
theory does provide resolution for some singularities, such as
orbifolds\orb, conifolds\andi\ and enhancons\enh. For spacelike
singularities, less has been achieved. For example, even the most familiar GR
singularity inside the Schwarzschild black hole has not yet been
understood.

Misner space can be embedded into string theory by adding 8
additional flat directions, and it is an exact solution of string
theory at least at tree-level \horowitz. The dynamics
of particles and strings in Misner universe were much explored in
the literature (see for example \niki\ \Berkz\ \Berka\ \Berkb\
\Berkc, for a good review see \pio). In particular, it was
realized in the above papers that winding strings are
pair-produced and they backreact on the geometry. Hence they may play important role in
the resolution of the singularity. Unfortunately, it is fair to
say that we still lack a sensible treatment of the backreactions.

Along another line, in the study of closed string tachyons\APS\
\min\ \sil, Misner space has reemerged as a valuable model\sm. By
imposing anti-periodic boundary conditions for fermions on the
spatial circle, one can get winding tachyons near the singularity
which can significantly deform the original geometry. It is
argued \sm\ that the spacetime near the spacelike singularity will
be replaced by a new phase of the tachyon condensate. In their case the
influence of the winding modes to the spacetime geometry is more significant and more
tractable. It is mainly this work \sm\ that motivates our following study.

We will use D-branes to probe the background geometry. D-branes are attractive here because
they can feel distances smaller than string scale \DKPS. For
Misner universe, the singularity is localized in sub-string region
in the time direction, so we will use D-instantons as probes.
Recently, D0 and D1-branes in Misner space were studied in \Jap, and it was found
that they are both unstable due to open string pair production and closed
string emission.

In fact, the usefulness of D-instantons has deeper reasons. It was
conjectured in \BFSS\ that the large $N$ limit of the
supersymmetric matrix quantum mechanics describing D0-branes
provides a holographic description of M-theory in the light cone
frame. In this model, known as BFSS matrix theory, all spatial
dimensions are dynamically generated while time is put in a prior.
Later in \IKKT, another matrix theory is proposed for IIB theory.
This so called IKKT matrix theory is a 0+0
dimensional theory, in contrast to the 0+1 dimensional BFSS theory.
Thus in this theory, both spatial and temporal dimensions
are generated dynamically. In fact, the IKKT action is essentially just the D-instanton action. Thus the D-instanton
action provides a holographic description of the full string theory. We will argue in this note that
this also happens for D-instantons in the Misner universe.

One of the advantages of the holographic
description is that backreaction can be taken into account more naturally, since the geometry and
objects in it are not treated seperatedly as in conventional theory
including perturbative string theory. And the drawback is that it is often tricky to get detailed
information from these matrices. And in this note we will encounter both the advantages and disadvantages.

The layout of this note is as follows. In section 2, we review some aspects
of the Misner geometry and properties of closed strings in it. We begin in section 3
the investigation of D-instanton physics. We derive the matrix action, and identify the
vacuum corresponding to the background geometry. And in section 4, we study how tachyons affect
 spacetime. We read from the tachyon
deformed D-instanton action the new moduli space, and thus the resulted
geometry. Finally in section 5, we promote our probe action to a second quantized framework, which
can be regarded as an Lorentzian orbifolded version of IKKT matrix theory.
We construct D-branes from matrices and study their properties, giving
evidence that the D-instanton action actually provides a holographic description for Misner universe.

Recent
explorations of other singularities include \mbb, \hhh, \yz.

Note added: After our paper was submitted to archive, we received
another paper \tai\ addressing similar problems.

\newsec{Misner universe: The geometry and the closed string story}
Misner universe is an orbifold of 1+1-dimensional Minkowski space
\eqn\mis{ds^2=-2dx^+dx^-}by the identification \eqn\orb{x^+\sim
e^{2\pi\gamma}x^+,\quad x^-\sim e^{-2\pi\gamma}x^- .} Coordinate
transformation \eqn\tran{x^+={T\over \sqrt{2}}
e^{\gamma\theta},\quad x^-={T\over \sqrt{2}} e^{-\gamma\theta}}can
be made to write Misner space as
\eqn\misa{ds^2=-dT^2+\gamma^2T^2d\theta^2,}with
$\theta\cong\theta+2\pi.$ It is easy to see from \misa\ that this
space-time contains two cosmological regions connected by a
space-like singularity.

There are generally two kinds of closed strings in Misner universe: twisted and untwisted.
Untwisted states include in particular the gravitons and their behaviors are particle-like. Their
wave functions can be obtained by superposing a plane wave in the parent Minkowski space
 with its images under the boost \orb, and is written as \niki\ \Berkb\
\eqn\untw{f_{j,m^2,s}(x^+,x^-)=\int^{\infty}_{-\infty}dv e^{ip^+X^-e^{-2\pi\gamma v}+ip^-X^+e^{2\pi\gamma v}+ivj+vs},}
with $j$ the boost momentum, $m$ the mass, and $s$ the $SO(1,1)$ spin in $R^{(1,1)}$.

Due to the orbifold projection \orb, new twisted sectors arise in Misner space
with strings satisfying \eqn\twist{X^{\pm}(\tau,\sigma+2\pi)=e^{\pm 2\pi\gamma w}X^{\pm}(\tau,\sigma),}
where the winding number $w$ is an integer. Many mysterieses of the Misner universe have origin from
these winding strings. It was shown in \Berka\ that
there exists a delta-function normalizable continuum of physical twisted states, which can be pair produced
in analogy with the Schwinger effect in an electric field. And evaluating the Bogolubov coefficients,
they showed that the transmission coefficient reads
\eqn\sch{q_4=e^{-\pi M^2 /2\nu}{{\cosh(\pi {\tilde M}^2/2\nu)}\over{|\sinh\pi j|}},} where
$\nu=-\gamma w$ is the product of the boost parameter and the winding number, and
\eqn\mass{M^2=\al^+_0\al^-_0+\al^-_0\al^+_0, \quad {\tilde M}^2=\at^+_0\at^-_0+\at^-_0\at^+_0}
with string zero modes $\al^{\pm}_0$ and $\at^{\pm}_0$, comes from the Virasoro conditions.

\newsec{D-instantons probing Misner universe}

We embed the geometry \mis\ \misa\ into string theory by adding
another 8 flat directions $Y^a, a=1, \dots, 8$, and then put $N$ D-instantons in this
geometry then go on to find the field theory describing their
behavior. We want to read from the modular space of the
D-instantons the background geometry, following the study of \APS\
\Berkooz . In this note we ignore the backreaction of these
D-instantons.

D-brane dynamics on the orbifolds were variously discussed in the
previous literature. We follow mainly Taylor's procedure \Taylor.
The open string degrees of freedom form a matrix theory. We focus
on the bosonic part, which are the embedding coordinates. Go to
the covering space \eqn\coor{(X^+,X^-)\in R^{1,1},\quad Y^a\in
R^8_{\perp},} and make the projection\orb, then each D-instanton
has infinitely many images, which can be captured by matrices of
infinitely many blocks. Each block is itself a $N\times N$ matrix.
The orbfold projection for these blocks reads \eqn\borb{\eqalign{
X^+_{i,j}&=e^{2\pi\gamma}X^+_{i-1,j-1},\cr
X^-_{i,j}&=e^{-2\pi\gamma}X^-_{i-1,j-1},\cr
Y^a_{i,j}&=Y^a_{i-1,j-1}.}} These matrices can be solved using the
following basis: \eqn\bet{(\beta^m_l)_{ij}=e^{2\pi il\gamma}
\delta_{i,j-m}.} Some of their communication relations will be
used in this note: \eqn\comm{\eqalign{[\beta^m_0,\beta^{m'}_0]&=0
 \cr [\beta^m_0,\beta^{m'}_1]&=(e^{2\pi m\gamma}-1)\beta^{m+m'}_1
\cr [\beta^m_0,\beta^{m'}_{-1}]&=(e^{-2\pi
m\gamma}-1)\beta^{m+m'}_{-1} \cr
[\beta^m_1,\beta^{m'}_{-1}]&=(e^{-2\pi m\gamma}-e^{2\pi
m'\gamma})\beta^{m+m'}_0.}} The solutions thus read
\eqn\solu{\eqalign{X^+&=\sum_{m\in Z} x^+_m\beta^m_1,\cr
X^-&=\sum_{m\in Z} x^-_m\beta^m_{-1},\cr Y^a&=\sum_{m\in Z}
y^a_m\beta^m_0. }}

The low energy effective action for the D-instantons can be
obtained from dimensional reduction of 10-d Super Yang-Mills, and
keep only the bosonic part, we get \eqn\acta{S={1\over{2 g^2
Z_0}}\sum_{\mu,\nu=0}^9 \Tr
([X^{\mu},X^{\nu}][X_{\mu},X_{\nu}]),}with coupling
$g^2={g_s\over\ap^2}$, where we eliminate factors of order 1;
$Z_0$ is the normalization factor which is formally trace of the
infinite dimensional unite matrix. The above
action is written in the Minkowski signature,
so there is an overall sign difference with IKKT\IKKT. Written in terms of the above
solution\solu, the action reads \eqn\actb{\eqalign{S&=-{1\over
g^2}\sum_{m+m'+n+n'=0} \Tr\biggm[x^+_m x^+_{m'}x^-_n
x^-_{n'}({e^{-2\pi m\gamma}-e^{2\pi n\gamma}})({e^{-2\pi
m'\gamma}-e^{2\pi n'\gamma}})\cr &+2 x^+_m x^-_{m'}y^a_n
y^a_{n'}({e^{2\pi n\gamma}-1})({e^{2\pi
n'\gamma}-1})e^{-2\pi(m+n)\gamma}\biggm].}} The above action has
many branches of vacuum. In the following, we will consider the
Higgs branch which corresponds to D-branes probing the Misner part
of the geometry, with the same coordinates in the other 8
directions. Thus we can eliminate the second term of the above
action.

The infinite summation in \actb\ indicates a "hidden"  dimension
with topology $S^1$, on which the Fourier coefficients of a real
scalar field can represent the modes in \solu. That is
\eqn\fou{\eqalign{x^+_m&=\int_0^{2\pi}{{d\sigma}\over{\sqrt{2\pi}}}X^+(\sigma)e^{-im\sigma},\cr
x^-_m&=\int_0^{2\pi}{{d\sigma}\over{\sqrt{2\pi}}}X^-(\sigma)e^{-im\sigma},
}}and the action \eqn\actc{\eqalign{S&=-{1\over
g^2}\int_0^{2\pi}{{d\sigma}\over{2\pi}}\Tr
\biggm([e^{i2\pi\gamma{d\over{d\sigma}}}X^+(\sigma)]X^-(\sigma)-
[e^{-i2\pi\gamma{d\over{d\sigma}}}X^-(\sigma)]X^+(\sigma)\biggm)^2\cr
&=-{1\over
g^2}\int_0^{2\pi}{{d\sigma}\over{2\pi}}\Tr\biggm[X^+(\sigma+i2\pi\gamma)X^-(\sigma)
-X^-(\sigma-i2\pi\gamma)X^+(\sigma)\biggm]^2}}is complemented by
the symmetry \eqn\sym{\eqalign{X^+(\sigma)&\rightarrow
e^{2\pi\gamma}X^+(\sigma),\cr X^-(\sigma)&\rightarrow
e^{-2\pi\gamma}X^-(\sigma),}}inherited from the orbifold
projection \orb.

Note that the system \actc\ \sym\ possesses a large variety of
vacua, which are simply solutions of the equation
\eqn\vva{X^+(\sigma+i2\pi \gamma)X^-(\sigma)-X^-(\sigma-i2\pi
\gamma)X^+(\sigma)=0.} Just to study
 the space-time geometry, we take all the D-instantons to coincide, and
 the matrices become ordinary functions. We can define \eqn\lll{F(\sigma)\equiv
X^+(\sigma+i2\pi\gamma)X^-(\sigma),}which appears repeatedly in this
note. And the vacuum condition reads
now\eqn\vvl{F(\sigma)=F(\sigma-i2\pi\gamma).} Note also that
$X^+,X^-$ are all defined on a circle, which says
\eqn\vvla{F(\sigma)=F(\sigma+2\pi).} For the problems in hand, we
expect $X^+(\sigma),X^-(\sigma)$ to have no poles in the $\sigma$
plane, so $F(\sigma)$ must be a constant.

Thus $X^+(\sigma+i2\pi\gamma)$ can be factorized as a real function
of $\sigma$ with periodicity $2\pi$ multiplied by a constant
\eqn\xaf{X^+(\sigma+i2\pi\gamma)=\alpha f(\sigma).} For real
functions $X^+(\sigma),f(\sigma)$ with periodicity $2\pi$, we can
expand them as \eqn\xandf{X^+(\sigma)=\sum^{+\infty}_{n=-\infty}c_n
e^{in\sigma}, \quad f(\sigma)=\sum^{+\infty}_{n=-\infty}f_n
e^{in\sigma},}with $c_{-n}=c_n^*$ and $f_{-n}=f_n^*$. Then eq.\xaf\
leads to \eqn\cgaf{c_n e^{-2\pi n\gamma}=\alpha f_n,} and thus
\eqn\cgafs{c_{-n} e^{2\pi n\gamma}=\alpha f_{-n},} or
\eqn\cgafa{c_n^* e^{2\pi n\gamma}=\alpha f_n^*,} and
\eqn\cgafb{c_n^* e^{-2\pi n\gamma}=\alpha^* f_n^*.} For any non
vanishing $c_n,f_n$, eq.\cgafa\ \cgafb\ require \eqn\cgafo{e^{4\pi
n\gamma}={\alpha\over{\alpha^*}},} which obviously can not be
satisfied for more than one value of $n$. And note that for $n$
non-zero, $c_n,f_n$ are paired with $c_{-n},f_{-n}$. So all
$c_n,f_n$ except $c_0,f_0$ must vanish, and thus $X^+(\sigma)$ must
be a constant, which subsequently forces $X^-(\sigma)$ also to be a
constant.

Taking into account the constraint \sym, we get a branch of the
moduli space (the Higgs branch)
 \eqn\moda{{\cal M}=\left\{X^+,X^-,Y^a\in R \biggm/\left\{\eqalign{
      X^+&\cong e^{2\pi\gamma}X^+ \cr
      X^-&\cong e^{-2\pi\gamma}X^-}
    \right\}\right\}}which is exactly the
original Misner universe.

To end this section, we remind the reader of some characteristics of
the action\actc. First, it is non-local. And the physical origin is
still mysterious to us. At first glance one may think winding modes
can cause such non-locality. But from the above calculation we see
that the effect of these twisted sectors is to induce the infinite
summation in eq.\actb\ and thus only leaving trace in the necessity
to use an integral in eq. \actc. In the null brane case \simon,
where there are similarly twisted sector contributions,  D-instanton
action is calculated in \Berkooz, which is also an integral but with the
integrand local. And we see that the non-locality is very peculiar
to Misner space whose singularity is spacelike.

 It was shown in \sethi\ by
Hashimoto and Sethi that the gauge theory on the D3-branes in
 the null
brane \simon\ background is noncommutative, thus also non-local.
What is interesting in their model is that they observe that upon
taking some decoupling limit, the noncommutative field theory
provides a holographic description of the corresponding
time-dependent closed string background (see also \ohta). Whether
some decoupling limit \ooo\ exists in our case is worth exploring.

Second, notice that the argument in the action \actc\ is
complexified, which is a peculiar property of some time-dependent
backgrounds. And it is also a crucial ingredient in our following
treatment of instability of Misner space and of the branes
therein. Complexified arguments also make appearance in the study
of other singularities (see for example \shenker, \hong).

\newsec{D-instantons probing tachyon deformed Misner universe}
We go on to deal with the case with winding string tachyon
condensates turned on \sm. Take anti-periodic boundary conditions
around the $\theta$ circle in \misa. In the regime
\eqn\tac{\gamma^2T^2\leq l_s^2, } some winding closed string modes
become tachyonic which signals the instability of the spacetime
itself. These modes grow and deform the spacetime. It was
speculated in \sm\ that the regime \tac\ will be replaced by a new
phase with all closed string excitations lifted.

 D-instantons feel the change in the geometry through
 its coupling to the metric. It was shown by Douglas and Moore in
  \dm\ that the leading effect of tachyons on the Euclidean orbifolds is to
induce a FI-type term in the D-brane potential. This effect comes
from the disk amplitude with one insertion of the twisted sector
tachyon field at the center and two open string vertex operators at
the boundary. With a detailed analysis of the full quiver gauge
theory, which provides a description for D-branes on the orbifolds, they combine the
FI term with the Born-Infeld action and the kinetic energies of the
hypermultiplets, and then integrate out the auxiliary D-fields in
the vectormultiplet, to find that the effect of the twisted sector
fields is to add a term in the complete square. In our case, we are
dealing with a Lorentzian orbifold which is more subtle than its Euclidean cousin.
 But to study the D-instanton theory, we can perform
a wick rotation to go to the Euclidean case, where the result of \dm\ will be consulted, and
 finally we get
schematically\eqn\actd{S=-{1\over
g^2}\int_0^{2\pi}{{d\sigma}\over{2\pi}}\biggm[X^+(\sigma+i2\pi\gamma)X^-(\sigma)
-X^-(\sigma-i2\pi\gamma)X^+(\sigma)-U(\sigma)\biggm]^2.} The
detailed form of $U(\sigma)$ is not important in the following
treatment where we require only the existence of such a non-zero
term. There may be some subtlty in the above wick rotation which deserves further clarification.
And the above D-instanton action can also be thought of as coming from a time-like
T-dual \hull\ of a more controllable system with D-particles on an Euclidean orbifold.

The vacuum condition becomes now\eqn\vvb{X^+(\sigma+i2\pi
\gamma)X^-(\sigma)-X^-(\sigma-i2\pi \gamma)X^+(\sigma)=U(\sigma).}
And the leading effect of the tachyons is to make the geometry
noncommutative \eqn\nca{[X^+(\sigma),X^-(\sigma)]=U(\sigma).} The
above approximation essentially sets $\gamma =0$. And we know that
the geometry corresponding to $\gamma=0$ is just the flat space
without any boost identification, so one may think this case can
not teach us much about Misner space. But we note that the term
$U(\sigma)$ encodes information peculiar to Misner space. From our
experiences for other better understood tachyons \sen\ \APS\ \min,
we can think this way: the nontrivial boost identification, plus
the anti-periodic boundary condition for fermions, first cooks
some closed string tachyons. Then these tachyons condense. For
these stages we cannot say anything new in the above formalism. We
intend only to explore how subsequently spacetime geometry is
modified by these tachyon condensates, taking into account the
fact that the tachyons couple to the metric. At this stage, the
process is driven by the tachyon source while the nontrivial boost
identification is not essential, thus we can use the above
approximation $\gamma=0$.

 To see the picture more clearly,
let's go to the $(T, \theta)$ frame. We can model the geometry by choosing
\eqn\ncb{e^{\gamma\theta}Te^{-\gamma\theta}=aT,}with $a$ some constant.
This makes a fuzzy cone, where the deviation of $a$ from
unity measures the fuzziness of the geometry. And the noncommutative relation \nca\ reads now
\eqn\ncc{T^2(a-{1\over a})=2U.}
We see from the above equation that in the two asymptotic regions $T\rightarrow\pm\infty$, $a$
goes to unity, thus $T$ commutes with $\theta$ and conventional geometric notion works well. But as $T$
goes to zero, $a$ deviates more and more from unity. Thus spacetime becomes more and more fuzzy.
At the origin $T=0$, $a$ diverges, and the conventional notion of geometry
breaks down totally. Thus the original spacelike singularity is removed.

\subsec{Comparison with McGreevy and Silverstein' s Nothing Phase}
It is also interesting to compare our result with that of \sm\
 (see also \stalk) which actually motivated our study. They employ
 perturbative string methods, working on the world sheet using
 techniques from Liouville theory. Here we will intend to propose
 a non-perturbative formulation of the theory, and the emerging picture is
in fact consistent with their work.$^{**}$ \footnote{}{**We give
literally different answer to the question: can time start or end
by turning on such closed string tachyons, where we employ
different interpretation of the question. They say yes \sm\ where
they mean conventional aspect of time breaks down in some region.
And we say no having in mind that information can still be
transferred from infinite past to infinite future. } They read
from
 their 1-loop partition function that the volume of the time
 direction is truncated to the region without closed string
 tachyon condensates, providing evidence for previously expected picture that closed
 string tachyon condensation lifts all closed string degrees of
 freedom, leaving behind a phase of "Nothing". In our formalism, we can say more about this
 "Nothing Phase". Although
 ordinary concepts of  spacetime break down, we can still model
 such region by some non-commutative geometry. Although closed
 string degrees of freedom cease to exist in such region, it is
 nevertheless possible to formulate the theory with open string degrees of
 freedom. And we expect the D-instanton matrix action \actd\ can serve
 this role. It seems that matrix models have the potential to say more about closed string tachyons,
 who are known as killers of closed string degrees of freedom, as open string tachyons did for open
string degress of freedom.

 Recently it was also found \mbb\ that near some null singularities, the usual supergravity and
even the perturbative string theory break down. Matrix degrees of
freedom become essential and the theory is more suitably described
by a Matrix string theory. Such non-abelian behavior seems
intrinsic for singularities.

\subsec{A model for big crunch/big bang transition}
The whole picture of the resulted spacetime after tachyon condensation is that of
 two asymptotically flat region, the infinite past and
infinite future, connected by some fuzzy cone. And although
conventional concept of time breaks down, there is still causal
connection between the infinite past and infinite future. This
fact is cosmologically attractive.

An alternative to inflation is proposed in \bang, where they considered the possibility
that the big bang singularity is not the termination of time, but a transition from
the contracting big crunch phase to the expanding big bang phase. The horizon
problem is nullified in this scenario, and other cosmological puzzles may also be solved in this new framework.
Unfortunately it is generally difficult
to get a controllable model for such a scenario.
From the above discussion, we see that the tachyon deformed Misner universe serves as a concrete model for
 such big crunch/big bang transition \bang.

\subsec{Remarks on space-time noncommutativity}
Note that the fuzzy cone condition \ncb\ is just a statement that space and time do not commute near the spacelike
singularity. This leads to the stringy
spacetime uncertainty relation, which was suggested by Yoneya and Li to
be a universal characterization of short distance structure for string and D-brane physics \li.
Here their idea is realized in a time-dependent background and we can compare with Hashimoto
and Sethi's realization \sethi\ of time-dependent space-space noncommutativity. In their case, the noncommutativity
can be traced back to the presence of background $B$ field, which is well understood \swnc. For Misner space the noncommutativity
all comes from the violent fluctuations of the geometry near the singularity, and this needs further study.

Space-time noncommutativity is also interesting for cosmology, since it leads to coupling
between inflation induced fluctuations and the background cosmology thus may produce transplackian effects.
 This subject is much explored in the literature \huang,
where it was pointed out that short distance dispersion relations may be modified, and non-Gaussianity,
anisotropism and the running of spectual index can be
explained. But usually for lack of concrete models, discussions are made generally. In the model of tachyon deformed Misner universe, more detailed questions can be asked.

\newsec{A possible holographic description of Misner universe}
Now we ask the question: how much information about the Misner
space is encoded in the action \actc ? In fact, the original
action \acta\ plus the fermionic part and a chemical potential
term proportional to $\Tr 1$ is proposed in \IKKT\ to provide a
constructive definition of definition of type IIB string theory.
This so called IKKT matrix theory, is interpreted in \tseyy\ as
the D-instanton counterpart of the D0-brane matrix theory of
BFSS\BFSS. Fundamental strings and Dp-branes \IKKT\ \tseyy\ \fdst\
can be constructed from such matrices, and long-distance
interaction potentials of BPS configurations computed from such
matrices match the supergravity results \tsey. We can extend this
matrix/string correspondence to the Misner case by making orbifold
projections \orb\ on both sides, where obviously the projection
commutes with the matrix/string mapping. So we propose that the
action \actc\ (plus its fermionic counterpart and possible
chemical potential term) provides a holographic description of the
Misner universe.

For these 0+0 (or here under orbifold projection, 1+0) dimensional
model, time is not put in a priori, but generated dynamically.
This may be the underlying reason why our description of change of
the structure of time is possible. And it also indicates that
these matrix models may have privilege in the description of
spacelike singularities and other time-dependent systems.

We go on to construct branes in Misner universe. The equation of
motion of action \acta\ is \eqn\eoma{g_{\mu\rho}g_{\nu\sigma}[
X^\nu,[X^\rho,X^\sigma] ]=0,} which in components are
\eqn\eomb{\eqalign{[X^-,[X^-,X^+]]-[Y^a,[X^-,Y^a]]&=0,\cr
[X^+,[X^+,X^-]]-[Y^a,[X^+,Y^a]]&=0,
\cr[X^+,[Y^a,X^-]]+[X^-,[Y^a,X^+]]&=0.}} As above we expand the
matrices in the $\beta$ basis \bet, and make the transformation
\fou, and choose $Y(\sigma)$ to constant. Thus we get the
classical solutions \eqn\eomc{\eqalign{\int
{{d\sigma}\over{2\pi}}\Tr
X^-(\sigma)\biggm(X^+&(\sigma+i4\pi\gamma)
X^-(\sigma+i2\pi\gamma)-X^-(\sigma)X^+(\sigma+i2\pi\gamma)\cr &-
X^+(\sigma+i2\pi\gamma)X^-(\sigma)
+X^-(\sigma-i2\pi\gamma)X^+(\sigma) \biggm)=0,\cr \int
{{d\sigma}\over{2\pi}}\Tr
X^+(\sigma)\biggm(X^-&(\sigma-i4\pi\gamma)X^+(\sigma-i2\pi\gamma)-
X^+(\sigma)X^-(\sigma-i2\pi\gamma)\cr
&-X^-(\sigma-i2\pi\gamma)X^+(\sigma)
+X^+(\sigma+i2\pi\gamma)X^-(\sigma) \biggm)=0.}} which are
unusually integral equations.

We can define \eqn\lll{L(\sigma)\equiv
X^+(\sigma+i2\pi\gamma)X^-(\sigma)-X^-(\sigma-i2\pi\gamma)X^+(\sigma),}which
appears repeatedly in this note. Note that $L(\sigma)=0$ is just
the vacuum \vva. Consider in the moduli space \moda\ a special
configuration
\eqn\dzero{\eqalign{X^+(\sigma)=X^-(\sigma)&=\pmatrix{t_1 & & &
\cr &  t_2 & & \cr & & \ddots & \cr & & &  t_N }  , \cr Y^a &=0,}}
where $t^{(i)}$ 's are constants. In the large $N$ limit, we see
from the corresponding classical trajectory \eqn\dzeroa{T=t,\quad
\theta=0,\quad Y^a=0 } with parameter $t$, that it is just a
D0-brane\IKKT. And the D-instanton action at this point of the
moduli space reproduces the D0-brane action of
 the BFSS matrix theory \IKKT.

It seems strange that D0-branes emerge this way from a D-instanton
matrix model which is directly related to the type IIB theory. In
flat space, these D0-branes are supersymmetric and stable.$^*$
\footnote{}{* Brane charges
are more subtle in the IKKT matrix theory than that of
BFSS \bss. Some proposals were made in \fdst\ for Dp-branes with p odd.
But for p even, (linear combinations of) the matrices commute with each other,
leaving no room for constructing central charges along the lines of \bss.
 } And in
Misner space they will also not decay for their seemingly wrong
dimension. The IKKT proposal \IKKT\ is that since type IIA string
theory is related to type IIB theory by T-duality, in some regions
of the type IIB moduli space, the type IIA theory can emerge as a
more suitable description. And the existence of these D0-branes
will be considered as manifestation of duality.

It is pointed out in \Jap\ that D0-branes in Misner universe are
actually unstable, they are subject to open string pair creation.
Since we regard our instanton action to be a second quantized
description of Misner universe, such phenomena should be
reproduced.

Let's make some small perturbation around the D0-brane \dzero
\eqn\dzerop{\eqalign{X^+(\sigma)&=\pmatrix{t_1+\delta t_1 & & &
\cr & t_2 & & \cr & & \ddots & \cr & & &  t_N },\cr
X^-(\sigma)&=\pmatrix{t_1 & & & \cr &  t_2 & & \cr & & \ddots &
\cr & & &  t_N }  , \cr Y^a &=0,}}with $\delta
t^{(1)}=\epsilon\sigma$. Now the action becomes
\eqn\actz{\eqalign{S&=-{1\over g^2}\int
{{d\sigma}\over{2\pi}}t_1^2[\delta t_1(\sigma+i2\pi\gamma)-\delta
t_1(\sigma)]^2\cr&=({{2\pi\gamma t_1}\over g})^2\epsilon^2.}}Note
the sign change above, which originates from the complexified
arguments in the integrand. And perturbations of other eigenvalues
give similar results.

To understand the above action, consider a quantum mechanical
system \eqn\qmm{S=\int dt \biggm[{1\over2}{\dot X}^2-U(X)
\biggm].} With $U(X)=-{1\over2}kX^2$ and $k>0$, it is just a
particle moving in an inverted harmonic potential. Seemingly the
particle can not stay static. It will roll down the potential.
When the potential term dominates the whole action, we go over to
the action \actz. And accordingly the D0-branes are unstable.
Worse still, the action is even not bounded from below, which
makes it impossible to define a first quantized vacuum. This fact
has already been noticed in \Berka\ in their study of perturbative
string theory of Misner universe.

Such kind of inverted harmonic potential also appears in $c=1$
matrix model, and there closed string emission from unstable
D0-branes is described by a matrix eigenvalue rolling down such a
potential \mvkms. In our formalism, description of such dynamical
processes is intrinsically subtle, where technically the
difficulty stems from the fact
 that we do not have
 kinetic terms for the matrix eigenvalues. But we can
understand from the path integral point of view that, the smaller
the Euclidean action $S_E=-S$, the more the configuration
contributes to the whole amplitude. And if we start with the
D0-brane \dzero, quantum fluctuations will generally destroy this
configuration, driving the system to more probable configurations
with larger $t_i$, making the brane effectively
 "roll down" the potential.
In the large $N$ limit, this corresponds to the phenomena that the
unstable D0-branes emit closed strings and/or open string pairs
\mvkms.

For the background geometry
\eqn\geoma{\eqalign{X^+(\sigma)&=x^+I_{N\times N},\cr
X^-(\sigma)&=x^-I_{N\times N},\cr Y^a(\sigma)&=y^a I_{N\times N,}
}}we can likely make a perturbation
\eqn\geomb{\eqalign{X^+(\sigma)&=\pmatrix{x^+ +\delta x^+ & & &
\cr & x^+ & & \cr & & \ddots & \cr & & &  x^+ },\cr
X^-(\sigma)&=x^-I_{N\times N},\cr Y^a(\sigma)&=y^a I_{N\times
N}.}} And similarly we get an inverted harmonic potential with
$\delta x^+$ a linear perturbation, thus the same instability
arises, which is consistent with what is found in perturbative
string theory in Misner space \Berka\ \Berkb\ which states that
the vacua is unstable while winding strings are pair produced as a
consequence of the singular geometry in analogy with the Schwinger
effect in an external electric field. This is a tunnelling
process, matching precisely our description via D-instantons. And
in this matrix framework, we can see that the instabilities of the
geometry and the branes have essentially the same origin. Both can
be interpreted as matrix eigenvalues "rolling down" a
unbounded-from-below potential.

Next let's discuss the D-strings. It is easy to see from \eomc\
that $L(\sigma)=constant$ is a classical solution. The Minkowski
limit $\gamma\rightarrow 0$ of $L(\sigma)$ is just the commutator
$[X^+(\sigma),X^-(\sigma)]$, and in this limit
$L(\sigma)=constant$ becomes the familiar result in matrix theory
\eqn\comg{[X^+(\sigma),X^-(\sigma)]=i{\cal F}^{+-}I_{N\times
N},}with ${\cal F}^{+-}$ some non-zero constant. And there in the
large $N$ limit, it represents D-strings \IKKT\ or some
non-marginal bound states of D-strings with D-instantons \tseyy.

Here the solution corresponding to a D-string is
\eqn\dstringa{\eqalign{X^+(\sigma)&={{L^+}\over{\sqrt{2\pi N}}}q,
\cr X^-(\sigma)&={{L^-}\over{\sqrt{2\pi N}}}p,
 \cr Y^a &=0 ,}} with $L^+$, $L^-$ some large enough compactification radius, and the $N\times N$ hermitian
matrices $0\leq q,p\leq \sqrt{2\pi N}$ satisfying
\eqn\ppqq{[q,p]=I_{N\times N},}which is obviously only valid at
large $N$. Note also the omitted $i$ in our convention in contrast
to usual notion.

These D-strings are also unstable \Jap, and the interpretation in
matrix theory is essentially the same as for D0-branes and the
geometry. We add some small perturbations, say change $q_{11}$ to
$q_{11}+\epsilon\sigma$, and the real part of the action becomes
now \eqn\actdd{S_{pert}=S_{D1}+({{2\pi\gamma p_{11}}\over
g})^2\epsilon^2,} leading to the "rolling" behavior of the matrix
elements and thus D-string's emitting open or closed strings.

The universality of the interpretation of instabilites of D0- and
D1-branes provides further evidence that D0-branes do not decay
for their "wrong dimensionality" and the region around \dzero\ has
a more suitable description as type IIA string theory.

Obviously more efforts are needed to figure out the details of the
string emission, such as the spectrum and emission rate which have
already been calculated in perturbative string theory \Berkb\
\Jap. The matrix formalism has the potential advantage to treat
more precisely the backreaction of the emitted strings as we have
exampled in section 4.

{\bf Acknowledgments}

I thank Bin Chen, Qing-Guo Huang, Miao Li and Peng Zhang for
valuable discussions. And in particular it's my pleasure to thank
Miao Li for permanent support and insightful comments at different
stages of this work.

\listrefs
\end